\documentstyle[titlepage,12pt]{article}
\def\singlespace {\smallskipamount=3.75pt plus1pt minus1pt
                  \medskipamount=7.5pt plus2pt minus2pt
                  \bigskipamount=15pt plus4pt minus4pt
                  \normalbaselineskip=15pt plus0pt minus0pt
                  \normallineskip=1pt
                  \normallineskiplimit=0pt
                  \jot=3.75pt
                  {\def\smallskip {\vskip\smallskipamount}}
                  {\def\medskip   {\vskip\medskipamount}}
                  {\def\bigskip   {\vskip\bigskipamount}}
                  {\setbox\strutbox=\hbox{\vrule
                    height10.5pt depth4.5pt width 0pt}}
                  \parskip 7.5pt
                  \normalbaselines}
\def\middlespace {\smallskipamount=5.625pt plus1.5pt minus1.5pt
                  \medskipamount=11.25pt plus3pt minus3pt
                  \bigskipamount=22.5pt plus6pt minus6pt
                  \normalbaselineskip=22.5pt plus0pt minus0pt
                  \normallineskip=1pt
                  \normallineskiplimit=0pt
                  \jot=5.625pt
                  {\def\smallskip {\vskip\smallskipamount}}
                  {\def\medskip   {\vskip\medskipamount}}
                  {\def\bigskip   {\vskip\bigskipamount}}
                  {\setbox\strutbox=\hbox{\vrule
                    height15.75pt depth6.75pt width 0pt}}
                  \parskip 11.25pt
                  \normalbaselines}
\def\doublespace {\smallskipamount=7.5pt plus2pt minus2pt
                  \medskipamount=15pt plus4pt minus4pt
                  \bigskipamount=30pt plus8pt minus8pt
                  \normalbaselineskip=30pt plus0pt minus0pt
                  \normallineskip=2pt
                  \normallineskiplimit=0pt
                  \jot=7.5pt
                  {\def\smallskip {\vskip\smallskipamount}}
                  {\def\medskip   {\vskip\medskipamount}}
                  {\def\bigskip   {\vskip\bigskipamount}}
                  {\setbox\strutbox=\hbox{\vrule
                    height21.0pt depth9.0pt width 0pt}}
                  \parskip 15.0pt
                  \normalbaselines}
\begin{document}
\newcount\sectionnumber
\sectionnumber=0

\title{Low Energy Grand Unification with SU(16)}
\author{Biswajoy Brahmachari \\
Theory Group, \\
Physical Research Laboratory, \\
Ahmedabad - 380009, \\
India.}
\vskip25pt
\date{March 8 1992}

\maketitle

\begin{abstract}
\doublespace 

We study the possibility of  achieving low unification scale in
a grand unification scheme based on the gauge group SU(16). Baryon
number symmetry being an explicit local gauge symmetry here
gauge boson mediated proton decay is absent . We present  in
detail a number of symmetry breaking patterns and the higgs field 
representations giving  rise to the  desired symmetry breakings 
and identify one chain giving low energy unification.
These higgs field representations are constructed in such a way 
that higgs mediated proton decay is absent. At the end we indicate 
the very rich low energy physics obtainable from this model which 
includes quark-lepton un-unified  symmetry and chiral color symmetry.
In brief some phenomenological implications are also studied.
\end{abstract}

\doublespace 
\section{Introduction}

Grand Unified Theories (GUTs) \cite{gutrev} offer the possibility of a
simple, but unified description of strong and electroweak
interactions. Typically in these models at some high energy
all the interactions arise out of a single Lagrangian which is
locally invariant under the gauge transformations of a single
simple Lie group called the unification group. A large spectrum
of GUTs are proposed in the literature which is broadly
classified by the unification group. In the minimal $SU(5)$ model
all interactions unite at a single step at an energy around
$10^{14} GeV$ therefore predicting  the  absence of any new physics
between the standard electroweak breaking scale($M_z$) and the
unification scale($M_U$) while the $SO(10)$ model admits
intermediate breakings of symmetry. On the other hand there are
models which are inspired by superstring theories, one of them
postulates the exceptional group $E_6$ as the unifying
group. This specific model predicts at least $12$ exotic fermions
on top of the $15$ standard fermions. All these theoretically
very attractive models have at least one common prediction namely
the decay of proton.

There has been a desperate search by the experimentalists to see
the signature of proton decay for the last decade and a
half. Contrary to the theoretical beliefs proton decay has not
been discovered. At present the lower limit of the half life of
proton is  a whooping $10^{32}$ years. The nonobservation of
proton decay has made all the above models a little less 
attractive.

At this juncture one interesting possibility is
that unification is achieved at a low energy scale which
means that the big desert of particle physics between the
electroweak scale and the unification scale becomes small but
other experimental constraints including that on the lifetime
of proton remains satisfied.  A grand unification scheme based
on $SU(16)$ as the unification group offers such a possibility.
It is worth noting here that due to the low unification scale
such models are free from the problems of grand unified monopoles
\cite{P.Pal}

Earlier works on low energy unification in GUTS considered $SU(15)$
as the unification group\cite{framp}. Here we extend the idea to the
left-right symmetric version of such a theory. We show that retaining
all the good features of $SU(15)$ we can also incorporate left-right
symmetry in intermediate stages. Unlike the $SU(15)$ GUT here Lepton
number is a local gauge symmetry which may survive to a low energy
scale. Right handed neutrino can be accommodated naturally as all the
fermions transform in the fundamental representation of $SU(16)$.

This paper is structured as following. In section $2$ and $4$  we give
the symmetry breaking chains and the Higgs field representations 
required for the breakings of symmetry. In section $3$ we present
some mathematical preliminaries useful for the calculation of
the mass scales. In section $5$ we calculate the mass scales
using the renormalization group equations.
In section $6$ we study some phenomenological
consequences of this model and in section $7$ we state the 
conclusions. In the appendix we give some group theoretic
essentials.

\section{Symmetry breaking and Higgs fields}

To achieve low energy unification we propose a number of symmetry
breaking chains. At the level of highest symmetry the theory is
invariant under the gauge group $SU(16)$. At and above this level
the coupling constant is that of the group $SU(16)$. With the
decrease in energy, the group goes through a number of symmetry
breaking phases, and the theory becomes least symmetric at the
present energies with the residual symmetry of $SU(3)$ color and
the symmetry of electromagnetic interactions. It is noteworthy
that the baryon number symmetry remains exact up to a very low
energy scale of a few $TeV$. This makes the proton stable in the
sense that the gauge boson mediated proton decay is
absent. Interestingly the completely un-unified symmetry group of
the quarks and leptons also appears at a low energy scale
together with the chiral color symmetry. The appearance of this
group  at a comparatively low scale makes this model worthy of
phenomenological studies{\cite{pheno}}.In this section we illustrate one chain
in some detail to draw attention to the underlying group theoretic
points. In section 4 we shall consider more possible chains before
finally going to calculate the symmetry breaking scales.     
\begin{center}
BREAKING CHAIN 1
\end{center}
Here  at first we give the breaking chains that can give rise to the 
standard model groups
$SU(3)_C$$\times$$SU(2)_L$$\times$$U(1)_Y$. We note here that
there can be in general a number of chains of descent to the 
standard model.

\begin{eqnarray}
SU(16)&{M_U \atop \longrightarrow}G[ SU(12) \times SU(4)^{l}] \nonumber \\
&{M_1 \atop \longrightarrow}G_1[ SU(6)_L \times SU(6)_R \times U(1)_B \times
SU(4)^{l}] \nonumber  \\
&{M_2 \atop \longrightarrow}G_2[ SU(3)_L \times SU(2)_L^{q}  \times
SU(6)_R \times U(1)_B\times SU(4)^{l}] \nonumber \\
&{M_3 \atop \longrightarrow}G_3[ SU(3)_L \times SU(2)_L^{q}\times SU(3)_R
\times U(1)_R^{q}\times U(1)_B\times SU(2)^{l}_L\times SU(2)^{l}_R
\times U(1)^{lep}]\nonumber\\
&{M_4 \atop \longrightarrow}G_4[ SU(3)_L \times SU(2)^{q}_L \times SU(2)^{l}_L
 \times SU(3)_R
\times U(1)_R\times U(1)_B \times U(1)^{l}]\nonumber\\
&{M_5 \atop \longrightarrow}G_5[ SU(3)_c \times SU(2)_L  \times
U(1)_B\times U(1)_h] \nonumber \\
&{M_6 \atop \longrightarrow}G_6[ SU(3)_c \times SU(2)_L  \times
U(1)_Y ]\nonumber \\  
&{M_z \atop \longrightarrow}G_7[ SU(3)_c \times U(1)_{em}] \nonumber 
\end{eqnarray}

Here the superscript q or l denotes that quarks or leptons have
nontrivial transformation law under these groups and the subscripts
L and R mean so for the left and right handed fermions. The subscript
c stands for the color gauge group of Q.C.D.

In a previous paper \cite{biswa1}we have shown that in $SU(15)$
GUT the effect of Higgs bosons play a significant role in the 
evolution of the coupling constants with increasing energy and 
hence on the values of the mass scales . This is due to the
presence of high dimensional Higgs fields required to obtain the
desired symmetry breaking pattern. The influence of the Higgs fields
on the evolution of coupling constants can be so serious that they
can alter the symmetry breaking pattern altogether. In $SU(16)$ GUT
The symmetry breaking pattern is very similar to that of its $SU(15)$
counterpart. So in $SU(16)$ or in $SU(15)$ GUT the Higgs effects must
be taken seriously. Here we shall consider the Higgs fields required 
to obtain the breaking chain and their contribution in the 
renormalization group equations in detail.

The Higgs structure is similar to that we proposed for $SU(15)$
GUT. We denote  $1^{n}$ as the totally antisymmetric nth rank tensor
and $1^{n}1^{m}$ as the representation which has m and n vertical boxes
in the first and second columns of its Young's table. For the transition
from the group $G_1$ to group $G_2$ the $G_2$ singlet component of the
Higgs field should acquire vacuum expectation value. Turning
to the specific case of $SU(16)$ we note that at the scale
$M_U$ the breaking can be achieved by giving the vacuum expectation
value to the $SU(12)$$\times$$SU(4)$ singlet component of $1^{4}$.
Using the exactly same procedure we see that the breaking at the
scale $M_1$ can be done by $1^{14}1$ which leaves $U(1)_B$ unbroken.
At the scale $M_2$ the breaking of $SU(6)_L$ to its special
maximal algebra requires a somewhat large dimensional Higgs field
representation. We use the $\bf{14144}$ dimensional Higgs field
$1^{14}1^{2}$ to break this group. As a passing comment we note
 here that this Higgs field will contribute significantly
to the beta functions of the renormalization group equations
and make its presence strongly felt in the determination of the
mass scales. The group $SU(4)^{l}$ can be broken by a Higgs field
which transforms as a $15$-plet under $SU(4)^{l}$ and which is contained
in $\bf255$ under $SU(16)$. At the stage $M_3$ the breaking of
$SU(6)_R$ to $SU(3)_R$$ \times$ $U(1)_R$ is a bit complicated. $\bf255$
breaks $SU(6)_R$ to $SU(3)$$ \times$ $SU(3)$$ \times $$U(1)_R$ and
 subsequently
the two $SU(2)_L$ groups of the quark and leptonic sectors
respectively are glued by $1^{14}1^{2}$. The breaking
of the lepton number local gauge symmetry $U(1)^{lep}$ can be
achieved by either $\bf{16}$ or the two index symmetric Higgs
field of dimension $\bf{136}$. In the first case it carries a
lepton number one unit and in the second case it carries that
of two units. We shall see that the choice of specific Higgs
field shall give interesting difference of physics in the
context of neutrino oscillations. At the scale $M_5$ the breaking
is done by the $1^{4}$ Higgs field which is $\bf{1820}$ dimensional.
The baryon number is broken by either $1^{5}$ or $1^{6}$. In
both the cases we get interesting physics.
As an example in the first case we get processes where baryon number 
changes by $3$ units and in the second case it changes by $2$ units. 
It is well-known that to give masses to the fermions vacuum expectation
value has to be given to the component $(1,2,-{1\over2})$ which is
contained in either $1^{2}$ or $11$. These Higgs Field
representations are summarized in table $1A$

Let us now turn our attention to the group theoretic transformation 
properties of the fermions under the different symmetry groups
in the symmetry breaking scheme. A minimal left-right symmetric
theory should have at least one right handed neutrino ($\nu_R$) on top of
the standard twelve quarks which includes three left handed
doublets and six right handed singlets under the weak interaction 
gauge group $SU(2)_L$ and three leptons namely
one left handed  doublet and one right handed singlet. At grand
unification energies and above this sixteen fermions should
transform under some representation of the unification group.
As a passing comment here we state that this requirement makes 
$SU(16)$ a very natural choice of the unification gauge group
which has a $16$ dimensional fundamental representation.
In the model the fermions transform under the fundamental
representation of $SU(16)$. Now as the energy becomes lower
the symmetry breakings occur and the transformation
properties of the fermions change with each symmetry breaking
taking place. In the following we summarize these transformation
properties. We use the notation that $(m,n)$ is a representation
which transforms under the semisimple group $SU(M)$$\times$$SU(N)$
as a m-plate under the former group and as a n-plate under the
the later group.
\begin{eqnarray}
SU(16){\longrightarrow}& 16  \nonumber \\
G{\longrightarrow}& (12,1)+(1,4) \nonumber  \\
G_{1}{\longrightarrow}&(1,\bar{6},n,1)
+(6,1,-n,1)+(1,1,0,4)\nonumber \\
G_{2}{\longrightarrow}&(1,1,\bar{6},n,1)+(3,2,1,-n,1)\nonumber\\
& +(1,1,1,0,4)\nonumber\\
G_{3}{\longrightarrow}&(1,1,\bar{3},p,n,1,1,0)
+(1,1,\bar{3},-p,n,1,1,0)\nonumber\\   \phantom{9}& +
(3,2,1,0,-n,1,1,0)
+(1,1,1,0,0,1,2,m)\nonumber \\ & +(1,1,1,0,0,2,1,-m)\nonumber \\
G_{4}{\longrightarrow}&(1,1,\bar{3},p,n,0)
+(1,1,\bar{3},-p,n,0)\nonumber \\ & +(3,2,1,0,-n,0)
+(1,1,1,0,0,-2{\sqrt{2\over3}} m)\nonumber \\
& +(1,2,1,0,0,{\sqrt{2\over3}} m)+(1,1,1,0,0,0) \nonumber \\
G_{5}{\longrightarrow}&(\bar{3},1,n,n)+(\bar{3},1,n,-n)+(3,2,-n,0)\nonumber
\\
& +(1,2,0,n)
+(1,1,0,-2n)+(1,1,0,0)\nonumber \\
G_{6}{\longrightarrow}&({\bar{3}},1,-{2\over{3}}K)+({\bar{3}},
1,{1\over{3}}K)\nonumber \\
& +(3,2,{1\over{6}}K)+(1,1,K)+(1,1,-{1\over{2}}K)\nonumber \\
& +(1,1,0) 
\end{eqnarray} 
Here the $U(1)$ normalisations are defined in terms of
\begin{eqnarray}
n & = & 1\over{2{\sqrt{6}}} \nonumber\\
m & = & 1\over{2{\sqrt{2}}} \nonumber\\
p & = & 1\over{2{\sqrt{3}}} \nonumber\\
K & = & \sqrt{3\over20}    \nonumber
\end{eqnarray}
We know that in the electroweak breaking scale $M_Z$ the
generators of electromagnetic symmetry group $U(1)_{em}$ arises out as
a linear combination of the generator of the $U(1)$ part of the weak
isospin group $SU(2)_L$ and that of the weak hypercharge $U(1)_Y$
by the following equation,
\begin{equation}
Q=T^{3}_L+Y
\end{equation}
Let us call this equation as the $U(1)$ $matching\phantom{1}condition$ 
at the scale $M_Z$. Similarly at the various symmetry breaking 
scales in the above breaking chain we have used different matching
conditions for the groups. These matching conditions are stated
below.

At the scale $M_4$ the lepton number symmetry  breaks
as the generator of $U(1)^{lep}$ and the diagonal generator
of $SU(2)^{l}_R$ mixes with each other in the following way
to generate the group $U(1)^{l}$,
\begin{equation}
Y^{l} = {\sqrt{1\over3}} T^{3}_{2^{l}_{R}} + {\sqrt{2\over3}} Y^{lep}
\end{equation}   

At the scale $M_5$ $U(1)_R$ and $U(1)^{l}$ breaks to make
$U(1)_h$.
\begin{equation}
Y_h  = {\sqrt{1\over2}} Y_R + {\sqrt{1\over2}} Y^{l}
\end{equation}

At  the scale $M_6$ ,baryon number cease to be  a local gauge 
symmetry and conventional hypercharge appears from
the linear combination of $U(1)_B$  and $U(1)_h$.
\begin{equation}
Y = -{\sqrt{1\over10}} Y_B - {\sqrt{9\over10}} Y_h
\end{equation}

\section{Mathematical Preliminaries}
In this section we briefly touch two more mathematically involved
topics . To begin with we note that the generators of $SU(16)$
and that of the standard model groups cannot be normalized in
the same way. We proceed further in the section by giving a short
discussion of the process of calculating the contribution of the 
Higgs fields to the beta functions.
Let us fix that all the generators of $SU(16)$ are normalized to
$2$. In that case at the standard model energies the generators of
$SU(3)_C$ and $SU(2)_L$ automatically becomes the generators of
$SU(16)$. In contrast the generators of $U(1)_Y$ are normalized
to $1\over2$. So in the renormalization group equations we have
to multiply the beta function corresponding to $U(1)_Y$ group by
the appropriate factor of $4$. Similarly it is easy to see that
all other $U(1)$ groups in the symmetry breaking chain has to
be multiplied by $4$. Turning to the non-Abelian groups it can
be checked that the group $SU(2)_L^{q}$ in all stages is normalized
to $3\over2$ hence to treat it at par with all other groups 
one has to multiply the beta function corresponding to this
by a factor of $4\over3$ . $SU(3)_L$ and $SU(3)_R$ in all the
stages are normalized to $1$ hence one finds the aforesaid
factor to be $2$. To complete the discussion on the normalization
factors we note that all other groups are normalized to $1\over2$
hence the relevant factor  is $4$

At this point let us turn our attention to the expression of
the beta function($b_i(N)$),for the group $SU(N)$
\begin{equation}
b(N)=-{1\over{(4\pi)}^{2}}\biggl\lbrack{{11\over3}N-{1\over6}T-{4\over3}n_f}
\biggr\rbrack  
\end{equation} 
For $U(1)$ groups N vanishes. Here  $n_f$ denotes the number of
families of fermions and $T(R)$ denotes the contribution of the
Higgs fields which transform nontrivialy  under the group under
consideration. To calculate $T$ we have followed the following
sum rule{\cite{slansky}}:

Suppose $R_i$  and $r_i$ ($i=1,2,..$) are different representations
of a group $SU(N)$,which when vectorically multiplied
satisfies the following relation.
\begin {equation}
R_1 \times R_2=\sum_{i=1}r_i \nonumber
\end{equation}
Also let for the representation of dimension r, the contribution
to the renormalization group equation is $T(R)$. Then,
\begin{equation}
T(R_1 \times R_2)=R_2 T(R_1)+R_1 T(R_2)=\sum_{i=1} T(r_i) \nonumber
\end{equation}
To use these equations one uses the following information to
start with

\begin{eqnarray}
T(N) & = & 1\over2 \nonumber\\
T(N^{2}-1) & = & N\nonumber\\
T \biggl\lbrack{{N(N-1)}\over2}\biggr\rbrack & = & {N-2}\over2
\nonumber\\
T \biggl\lbrack{{N(N+1)}\over2}\biggr\rbrack & = & {N+2}\over2
\nonumber\\
T(1) & = & 0 \nonumber
\end{eqnarray}

As an example consider $3$ and $\bar{3}$ representations of
$SU(3)$. When vectorically multiplied they give
\begin {equation}
3\times{\bar{3}}=1 + 8
\end{equation}
so using the sumrule 
\begin{equation}
T(8)= 3 T(3) + 3 T(\bar{3}) -T(1) = 3 \nonumber
\end{equation}

\section{Other paths to the Standard Model}
We have already noted that there can be a number of paths to the
standard model groups starting from the unification group $SU(16)$.
 Let us consider here two typical chains of descent. In the
first case (chain 2) here we shall break the $U(1)$ groups
as low as possible. It is in a sense one extreme case as the beta
function coefficients for the $U(1)$ groups are very small in magnitude
compared to that of the other groups ( the eigenvalue of  the 
casimir operator vanish ). 

\begin{center}
BREAKING CHAIN 2
\end{center}

\begin{eqnarray}
SU(16)&{M_U \atop \longrightarrow}G[ SU(12) \times SU(4)^{l}] \nonumber \\
&{M_1 \atop \longrightarrow}G_1[ SU(6)_L \times SU(6)_R \times U(1)_B \times
SU(4)^{l}] \nonumber  \\
&{M_2 \atop \longrightarrow}G_2[ SU(3)_L \times SU(2)_L^{q}  \times
SU(6)_R \times U(1)_B\times SU(4)^{l}] \nonumber \\
&{M_3 \atop \longrightarrow}G_3[ SU(3)_L \times SU(2)_L^{q}\times SU(3)_R
\times U(1)_R^{q}\times U(1)_B\times SU(2)^{l}_L\times SU(2)^{l}_R
\times U(1)^{lep}]\nonumber\\
&{M_4 \atop \longrightarrow}G_4[ SU(3)_c \times SU(2)^{q}_L\times U(1)^{q}_R
\times U(1)_B \times SU(2)^{l}_L\times SU(2)^{l}_R\times U(1)^{lep}]\nonumber\\
&{M_5 \atop \longrightarrow}G_5[ SU(3)_c \times SU(2)^{q}_L  \times
SU(2)^{l}_L\times U(1)^{q}_R\times U(1)_B\times U(1)^{l}_R\times
 U(1)^{lep}] 
\nonumber \\
&{M_6 \atop \longrightarrow}G_6[ SU(3)_c \times SU(2)_L  \times
U(1)_Y ]\nonumber \\  
&{M_z \atop \longrightarrow}G_7[ SU(3)_c \times U(1)_{em}] \nonumber 
\end{eqnarray}

It is to note that all the $U(1)$ groups remain ununified till
the scale $M_6$. At the scale they merge together to give the
familiar hypercharge of the standard model. At the scale $M_6$
the matching condition is
\begin{equation}
Y =-{\sqrt{2\over20}}Y_{B}-{\sqrt{9\over20}}Y^{q}_{R}
-{\sqrt{3\over20}}Y^{l}_R-{\sqrt{6\over20}} Y^{lep}
\end{equation}   

Another interesting possibility is to break $SU(16)$ via the left-right
symmetric group of Pati and Salam{\cite{pasa}}.The low energy Phenomenology
of the Pati-Salam group is widely studied. So it will be
interesting to see how low the intermediate scales can come down
to so we can make some concrete predictions of the model in view
of the on coming experiments. Hence
in the second chain that is discussed here (chain 3) the left-right symmetric
group $SU(3)_{c}\times SU(2)_L\times SU(2)_R\times U(1)_{B-L}$ will be
kept as low as possible  
\begin{center}
BREAKING CHAIN 3
\end{center}
\begin{eqnarray}
SU(16)&{M_U \atop \longrightarrow}G[ SU(12) \times SU(4)^{l}] \nonumber \\
&{M_1 \atop \longrightarrow}G_1[ SU(6)_L \times SU(6)_R \times U(1)_B \times
SU(4)^{l}] \nonumber  \\
&{M_2 \atop \longrightarrow}G_2[ SU(3)_L \times SU(2)_L^{q}  \times
SU(6)_R \times U(1)_B\times SU(4)^{l}] \nonumber \\
&{M_3 \atop \longrightarrow}G_3[ SU(3)_L \times SU(2)^{q}_L
\times SU(3)_R \times SU(2)^{q}_R
\times U(1)_B \times SU(4)_L\nonumber\\
&{M_4 \atop \longrightarrow}G_4[ SU(3)_c \times SU(2)^{q}_L\times SU(2)^{q}_R
\times U(1)_B \times SU(2)^{l}_L\times SU(2)^{l}_R\times U(1)^{lep}]\nonumber\\
&{M_5 \atop \longrightarrow}G_5[ SU(3)_c \times SU(2)^{q+l}_L  \times
SU(2)^{q+l}_R\times U(1)_{B-L}]\nonumber \\
&{M_6 \atop \longrightarrow}G_6[ SU(3)_c \times SU(2)_L  \times
U(1)_Y ]\nonumber \\  
&{M_z \atop \longrightarrow}G_7[ SU(3)_c \times U(1)_{em}] \nonumber 
\end{eqnarray}
We notice that $U(1)_{B-L}$ group is formed at the scale $M_5$
when baryon number symmetry and the lepton number symmetry is
broken together at the same scale. The matching condition is
\begin{equation}
Y_{B-L} = \sqrt{1\over4} Y_{B}-{\sqrt{3\over4}} Y^{lep}
\end{equation}
At the scale $M_6$  the generator of the group $U(1)_{B-L}$ and
the diagonal generator of the right handed $SU(2)$ group form a
linear combination to generate the  conventional hypercharge.
\begin{equation}
Y = -{\sqrt{9\over10}} T^{3}_{R}-{\sqrt{1\over10}} Y_{B-L}
\end{equation}
Applying exactly the similar principles that we used for
calculating the Higgs structure for the first breaking chain,
We can calculate the Higgs fields required to break $SU(16)$ in
the fashion of chain 1. The essential difference in chain 2 is
that the breaking of $SU(2)^{l}_R$ to $U(1)^{l}_R$ which can be
conveniently done by $\bf{255}$ which has a component (1,15)
under G  and at the scale $M_{6}$ the four $U(1)$ groups are
glued by a combination of Higgs fields $1^{4}$ and $1^{3}$.
The breaking chain 3 is much more symmetric and simpler too. The
Higgs fields that we require to break the chain is also less
complicated. To break the left right symmetry group we need the
Higgs fields
$(1,1,3,-{\sqrt{3\over8}})$, $(1,2,2,0)$, $(1,2,2,0)$
and $(1,1,3,\sqrt{3\over8})$ which can be easily embeded in 
the group $G$ in the representations (143,6) ,(1,15) and (78,1)
and hence in $SU(16)$. The details of the Higgs fields required
for the chains 2 and 3 are given in table 2A and table 3A.

\section{Mass scales}
To evaluate the mass scales we use the standard procedure of
evolving the couplings with energy. The energy dependence of the
couplings with energy{\cite{gqw}}. The energy dependence of the couplings
are completely determined by the particle content of the theory
and their couplings inside the loop diagrams of the guage
bosons. This is expressed by the renormalization group equation.
The one-loop RG equation is given by the following equation. 
\begin{equation}
\mu{d\over{d\mu}}\alpha(\mu) = 2b{\phantom{1}}\alpha^{2}(\mu) \nonumber\\
\end{equation}
{\rm where}
\begin{equation}
\alpha ={ g^{2}\over{4\pi}}
\end{equation}
the beta function coefficients are already defined. 
Now, using the above informations and
the matching conditions given with each symmetry breaking chain
one can relate the $SU(16)$ coupling constant $\alpha_{SU(16)}$
with the standard  model couplings $\alpha_{3c}$,$\alpha_{2L}$ and
$\alpha_{1Y}$ at the scale of the mass of the Z particle $M_Z$. At
this point let us remember that there are three quark doublets
and one leptonic doublet under the group $SU(2)_{L}$ in the
standard model hence in the evolution of coupling $\alpha_{2L}$
the quark and leptonic groups $SU(2)^{q}_L$ and $SU(2)^{l}_L$ do
not contribute equally to the standard model group $SU(2)_L$
instead they contribute with a relative factor 3   
\begin{center}

CHAIN 1
\end{center}
\begin{eqnarray}
g_{3c}^{-2}(M_z) & = &  g^{-2}_{SU(16)} (M_U) +\nonumber\\
& & 2b_{12}M_{U1}+(b_{6L}+b_{6R})M_{12}+(b_{3L}+b_{6R})M_{23}+\nonumber\\
& & (b_{3L}+b_{3R})M_{34}+ (b_{3L}+b_{3R})M_{45}+2b_{3c}M_{56}+2b_{3c}M_{6z}\nonumber\\
g_{2L}^{-2}(M_z) & = &  g^{-2}_{SU(16)} (M_U) +\nonumber\\ 
& & ({3 \over 2}b_{12} +{1 \over 2}b_{4}^l) M_{U1}+
({3 \over 2}b_{6L} +{1 \over 2}b_{4}^l) M_{12}+\nonumber\\
& & ({3 \over 2}b_{2L}^q +{1 \over 2}b_{4}^l) M_{23}
+({3 \over 2}b_{2L}^q +{1 \over 2}b_{2L}^l) M_{34}+
({3 \over 2}b_{2L}^q +{1 \over 2}b_{2L}^l)M_{45}+\nonumber\\
& &   2b_{2L}M_{56}+2b_{2L}M_{6z}
\nonumber \\ 
 g_{1Y}^{-2}(M_z) & = &   g^{-2}_{SU(16)} (M_U)+\nonumber\\ 
& & ({11 \over {10}} b_{12} + {9 \over {10}} b_{4}^{l})M_{U1}
+({9\over{10}}b_{6R}+{1\over{5}}b_{1B}+{9\over{10}}b_{4}^{l})M_{12}+\nonumber\\
& & ({9\over{10}}b_{6R}+{1\over{5}}b_{1B}+{9\over{10}}b_{4}^{l})M_{23}
+({9\over{10}}b_{1R}+{1\over{5}}b_{1B}+{6\over{10}}b_{1}^{lep}+{3\over{10}}b_{2R}^{l}+)
M_{34}+\nonumber\\
& & ({9\over{10}}b^{q}_{1R}+{1\over{5}}b_{1B}+{9\over{10}}b^{l}_{1})
M_{45}+\nonumber\\
& & ({9\over{5}}b_{1h}+{1\over{5}}b_{1B})M_{56}+
2b_{1Y}M_{6z} 
\end{eqnarray}

\begin{center}
CHAIN 2
\end{center}
\begin{eqnarray}
g_{3c}^{-2}(M_z) & = &  g^{-2}_{SU(16)} (M_U) +\nonumber\\
& & 2b_{12}M_{U1}+(b_{6L}+b_{6R})M_{12}+(b_{3L}+b_{6R})M_{23}+\nonumber\\
& & (b_{3L}+b_{3R})M_{34}+2b_{3c}M_{45}+2b_{3c}M_{56}+2b_{3c}M_{6z}\nonumber\\
g_{2L}^{-2}(M_z) & = &  g^{-2}_{SU(16)} (M_U) +\nonumber\\ 
& & ({3 \over 2}b_{12} +{1 \over 2}b_{4}^l) M_{U1}+
({3 \over 2}b_{6L} +{1 \over 2}b_{4}^l) M_{12}+\nonumber\\
& & ({3 \over 2}b_{2L}^q +{1 \over 2}b_{4}^l) M_{23}
+({3 \over 2}b_{2L}^q +{1 \over 2}b_{2L}^l) M_{34}+\nonumber\\
& &({3 \over 2}b_{2L}^q +{1 \over 2}b_{2L}^l) M_{45} +
({3 \over 2}b_{2L}^q +{1 \over 2}b_{2L}^l) M_{56}+2b_{2L}M_{6z}
\nonumber \\ 
 g_{1Y}^{-2}(M_z) & = &   g^{-2}_{SU(16)} (M_U)+\nonumber\\ 
& & ({11 \over {10}} b_{12} + {9 \over {10}} b_{4}^{l})M_{U1}
+({9\over{10}}b_{6R}+{1\over{5}}b_{1B}+{9\over{10}}b_{4}^{l})M_{12}+\nonumber\\
& & ({9\over{10}}b_{6R}+{1\over{5}}b_{1B}+{9\over{10}}b_{4}^{l})M_{23}
+({9\over{10}}b^{q}_{1R}+{1\over{5}}b_{1B}+{3\over{10}}b_{2R}^{l}+{6\over{10}
}b_{1}^{lep})M_{34}+\nonumber\\
& & ({9\over{10}}b^{q}_{1R}+{1\over{5}}b_{1B}+
{3\over{10}}b_{2R}^{l}+{6\over{10}
}b_{1}^{lep})M_{45}+\nonumber\\
& & ({9\over{10}}b_{1R}+{1\over{5}}b_{1B}+{3\over{10}}b_{1R}^{l}+{6\over{10}
}b_{1}^{lep})M_{56}
+2b_{1Y}M_{6z} 
\end{eqnarray}

\begin{center}
CHAIN 3
\end{center}
\begin{eqnarray}
g_{3c}^{-2}(M_z) & = &  g^{-2}_{SU(16)} (M_U) +\nonumber\\
& & 2b_{12}M_{U1}+(b_{6L}+b_{6R})M_{12}+(b_{3L}+b_{6R})M_{23}+\nonumber\\
& & (b_{3L}+b_{3R})M_{34}+2b_{3c}M_{45}+2b_{3c}M_{56}+2b_{3c}m_{6z}\nonumber\\
g_{2L}^{-2}(M_z) & = &  g^{-2}_{SU(16)} (M_U) +\nonumber\\ 
& & ({3 \over 2}b_{12} +{1 \over 2}b_{4}^l) M_{U1}+
({3 \over 2}b_{6L} +{1 \over 2}b_{4}^l) M_{12}+\nonumber\\
& & ({3 \over 2}b_{2L}^q +{1 \over 2}b_{4}^l) M_{23}
+({3 \over 2}b_{2L}^q +{1 \over 2}b_{4}^l) M_{34}+
({3 \over 2}b_{2L}^q +{1 \over 2}b_{2L}^l) M_{45}+\nonumber\\
& &   2b^{q+l}_{2L}M_{56}+2b_{2L}M_{6z}
\nonumber \\ 
 g_{1Y}^{-2}(M_z) & = &   g^{-2}_{SU(16)} (M_U)+\nonumber\\ 
& & ({11 \over {10}} b_{12} + {9 \over {10}} b_{4}^{l})M_{U1}
+({9\over{10}}b_{6R}+{1\over{5}}b_{1B}+{9\over{10}}b_{4}^{l})M_{12}+\nonumber\\
& & ({9\over{10}}b_{6R}+{1\over{5}}b_{1B}+{9\over{10}}b_{4}^{l})M_{23}
+({9\over{10}}b^{q}_{2R}+{1\over{5}}b_{1B}+{9\over{10}}b_{4}^{l})
M_{34}+\nonumber\\
& & ({9\over{10}}b^{q}_{2R}+{1\over{5}}b_{1B}+{3\over{10}}b_{2R}^{l}+{6\over{10}
}b_{1}^{lep})
M_{45}+\nonumber\\
& & ({9\over{5}}b^{q+l}_{2R}+{1\over{5}}b_{1(B-L)})M_{56}+
2b_{1Y}M_{6z} 
\end{eqnarray}

Here $M_{ij}$ is defined as $ln({M_{i}\over{M_{j}}})$
As a comment we note that generally one would expect that the
coefficients of $b^{q+l}_{2R}$ and $b_{1(B-L)}$ to be ${6\over5}$
and $4\over5$ respectively as it appears in the $SO(10)$ model. It
is worth noting that these factors are dependent on the
normalization of the $U(1)$ part of $SU(2)^{q+l}_{R}$ and that
of the group $U(1)_{B-L}$. This point will be elaborated further
in the appendix.

To calculate the mass scales we also have to know the numerical
values of the beta function coefficients. To know them one has
to know the contribution of the Higgs scalars to the beta
functions $(T)$. In the following three tables we give these
values.

\begin{table}[htb]
\begin{center}
\[
\begin{array}{|l|l|l|l|l|l|l|}
\hline
G&G_{1}&G_{2}&G_{3}&G_{4}&G_{5}&G_{6}\\
\hline
[12]=1492&[6_L]=69&[3_{L}]=42&[3_{L}]=15&
[3_L]=9&[3_c]=0&[3_c]=0 \cr
[4^{l}]=293&[6_R]=93&[2^{q}_L]=63&[2^{q}_L]=22.5&
[2^{q}_L]=13.5&[2_L]=0.5&[2_L]=0.5\cr
&[1_B]=45&[6_R]=93&[3_R]=15&
[3_R]=9&[1_B]=.375&[1_Y]=.075\cr
&[4^{l}]=63&[1_B]=45&[1^{q}_R]=7.58&
[1^{q}_R]=3.16&[1_{h}]=.083&\cr
&&[4^{l}]=63&[1_B]=.375&[1_{B}]=.375&&\cr
&&&[2^{l}_{L}]=18&[2^{l}_{L}]=9&&\cr
&&&[1^{lep}]=2&[1_l]=3.16&&\cr
\hline
\end{array}
\]
\end{center}
\caption{Contributions of the Higgs scalar to the R-G equation
at various energy scales in chain1.}
\end{table}
\begin{table}[htb]
\begin{center}
\[
\begin{array}{|l|l|l|l|l|l||l|}
\hline
G&G_{1}&G_{2}&G_{3}&G_{4}&G_{5}&G_{6}\\
\hline
[12]=1459&[6_L]=111&[3_{L}]=103&[3_{L}]=36&
[3_c]=0&[3_c]=0&[3_c]=0 \cr
[4^{l}]=109&[6_R]=120&[2^{q}_L]=99&[2^{q}_L]=31.5&
[2^{q}_L]=3.5&[2^{q}_L]=2.5&[2_L]=0.5\cr
&[1_B]=7.5&[6_R]=102&[3_R]=36&
[1^{q}_R]=1.25&[2^{l}_{L}]=2&[1_Y]=.075\cr
&[4^{l}]=62&[1_B]=7.5&[1^{q}_R]=7.58&
[1_B]=.375&[1^{q}_R]=.583&\cr
&&[4^{l}]=78&[1_B]=.375&[2^{l}_{L}]=3&[1_{B}]=.375&\cr
&&&[2^{l}_{L}]=27&[2^{l}_{R}]=4&[1^{l}_{R}]=1&\cr
&&&[1^{l}_{R}]=20&[1^{lep}]=.5&[1^{lep}]=12.5&\cr
&&&[1^{lep}=4.5&&&\cr
\hline
\end{array}
\]
\end{center}
\caption{Contributions of the Higgs scalar to the R-G equation
at various energy scales in chain2.}
\end{table}
\begin{table}[htb]
\begin{center}
\[
\begin{array}{|l|l|l|l|l|l||l|}
\hline
G&G_{1}&G_{2}&G_{3}&G_{4}&G_{5}&G_{6}\\
\hline
[12]=1831&[6_L]=171&[3_{L}]=79.5&[3_{L}]=57&
[3_c]=0&[3_c]=0&[3_c]=0 \cr
[4^{l}]=646&[6_R]=105&[2^{q}_L]=71&[2^{q}_L]=77&
[2^{q}_L]=8&[2^{q+l}_L]=4&[2_L]=1.5\cr
&[1_B]=105&[6_R]=117&[3_R]=54&
[2^{q}_R]=7&[2^{q+l}_R]=4&[1_Y]=.225\cr
&[4^{l}]=149&[1_B]=3&[2^{q}_R]=75&
[1_B]=1.5&[1_{(B-L)}]=2.25&\cr
&&[4^{l}]=89&[1_B]=3&[2^{l}_{L}]=6&&\cr
&&&[4^{l}]=51&[2^{l}_{R}]=5&&\cr
&&&&[1^{lep}]=3.5&&\cr
\hline
\end{array}
\]
\end{center}
\caption{Contributions of the Higgs scalar to the R-G equation
at various energy scales in chain3.}
\end{table}

Now with the quantities $ g_{1Y}^{-2}(M_z)$ $ g_{2L}^{-2}(M_z)$ and
$ g_{3c}^{-2}(M_z)$ at hand one can construct two different
linear combinations with them to form the experimentally
measured quantities at the energy scale $M_{z}$.It easy to check
that the following relations hold between them.
\begin{eqnarray}
Sin^{2}(\theta_{w}) & = & {3\over{8}}-{5\over{8}}e^{2}(g^{-2}_{1Y}-g^{-2}_{2L})
\nonumber\\
1-{8\over3}{\alpha\over\alpha_{s}}& = &
e^{2}(g^{-2}_{2L}+{5\over3} g^{-2}_{1Y} -{8\over3}g^{-2}_{3c})
\end{eqnarray}
From the present experimental measurements at LEP the value of 
$Sin^{2}(\theta_{w})$ and $\alpha_{s}$ has been very accurately
measured. We use for our purpose the following values{\cite{amaldi}} of them
and the $U(1)$ coupling $\alpha$ at the scale $M_{z}$
\begin{eqnarray}
Sin^{2}(\theta_{w}) & = & .2336\pm 0.0018 \nonumber\\
\alpha_{s} & = & .108\pm 0.005 \nonumber\\
\alpha & = & 1\over{128.8}
\end{eqnarray}
Having these informations at hand one can straightaway go to
calculate the mass scales of symmetry breaking.

Let us discuss the calculation  of the first chain in some
detail. Let us now assume that $M_{4}=M_{3}=M_{A}$. This means
that the groups $SU(6)_{L}$  $SU(6)_{R}$ and $SU(4)^{l}$ happens
to break at the same scale. Similarly let us also assume that 
$M_{4}=M_{5}=M_{B}$. Now using the values of the $T(R)$s 
 table2  and solving for $M_{U1}$
and $M_{B6}$ in terms of the other variables one gets,
\begin{eqnarray}
 M_{U1} & = & -.28-.10 M_{1A}-.10 M_{6z}+.04 M_{AB}   \nonumber\\
 M_{B6} & = &  19.80- 4.81 M_{1A}-2.93 M_{6z}- .21M_{AB}
\end{eqnarray}
As the symmetry breaking at $M_U$ occurs before it happens at
$M_1$ $M_{U1}$ is at least positive. So from the first equation
one infers that for a specific set of values of the other
parameters in the right-hand side there is a minimum value to
$M_{AB}$.Varying the parameters of the equations one gets the
following subset of the solution set allowed by the
equations. Taking $M_{z}$ to be around 91 GeV one can also
calculate the unification scale and the scale$M_{6}$ where
the completely un-unified symmetry of the quarks and leptons and
the chiral color symmetry is broken. We note that as the
parameter $M_{AB}$ increases i.e. as the separation between the
scale $M_{A}$ and the scale $M_{B}$ increases the scale $M_{B}$
comes down.
\begin{table}[htb]
\begin{center}
\begin{tabular}{|c||c||c||c||c||c||c|}
\hline
$M_{AB}$&$M_{1A}$&$M_{6z}$&$M_{B6}$&$M_{U1}$&$M_{B}$&$M_{U}$\\
\hline
7&0&0&18.4&0&$10^{9}$&$10^{12}$	\\
9.5&1&0&12.9&0&$10^{8}$&$10^{12}$ \\ 
10.75&1.5&0&10.3&0&$10^{7}$&$10^{11}$	\\
12&2&0&8.7&0&$10^{6}$&$10^{11}$\\
14.5&3&0&2.3&0&$10^{4}$&$10^{11}$	\\
\hline
\end{tabular}
\end{center}
\caption{Mass scales from chain1.}
\end{table}
Exactly in a similar way let us see the breaking scales that we
may get from the solution of the chain2. Here we
keep the $U(1)$ groups as low as possible in the hope that it
will give rise to distinct phenomenology at the low energy.
To begin with let us keep $M_{2}$=$M_{3}$=$M_{A}$ and
$M_{4}$=$M_{5}$=$M_{B}$ .The solutions are  
\begin{eqnarray}
 M_{U1} & = & -.70-.01 M_{1A}-.02 M_{6z}+.01 M_{AB}   \nonumber\\
 M_{B6} & = &  5.35-.44M_{1A}-.78M_{6z}- .81M_{AB}
\end{eqnarray}
To keep $M_{U1}$ positive we have to have $M_{AB}$ larger than 
70.This pushes the unification scale beyond the Planck scale and
hence makes the breaking chain uninteresting.

The third option that we have considered here is to come to the
low energy groups via the left-right symmetric  Pati-Salam group
in the chain3. The solutions in this case are
\begin{eqnarray}
 M_{U1} & = & -.99+.13 M_{1A}+.01 M_{56}+.02 M_{A5}   \nonumber\\
 M_{6z} & = &  6.84+2.64 M_{1A}-.22 M_{56}- .1 M_{A5}
\end{eqnarray}
The solution set of these equations are interesting though low
energy unification is not possible here. Let us at first state a
sample solution set.
\begin{table}[htb]
\begin{center}
\begin{tabular}{|c||c||c||c||c||c||c|}
\hline
$M_{1A}$&$M_{A5}$&$M_{56}$&$M_{U1}$&$M_{6z}$&$M_{6}$&$M_{U}$\\
\hline
7.61&0&0&0&26.92&$10^{13}$&$10^{17}$	\\
6.84&5&0&0&24.89&$10^{12}$&$10^{18}$ \\ 
6.07&10&0&0&21.86&$10^{11}$&$10^{18}$	\\
5.30&15&0&0&19.35&$10^{10}$&$10^{19}$\\
4.53&20&0&0&16.79&$10^{9}$&$10^{19}$	\\
\hline
\end{tabular}
\end{center}
\caption{Mass scales from chain3.}
\end{table}
This is obvious from the equations that to keep $M_{U1}$
positive one need a rather large value of $M_{1A}$ which on the
other hand pushes $M_{6z}$ up. The minimum value of $M_{1A}$ is
around 7.6 which gives the minimum value of the unification
scale which is around $10^{17}$ GeV. In a previous paper{\cite{mlr}} we
have shown that with the precisely measured value of
$Sin^{2}(\theta_{w})$ that is available now left-right symmetry
at the low energy coming from a grand unified scenario is ruled
out. This analysis comes as a confirmation of that result and it
shows that even having a number of parameters that we have in
the form of a number of breaking stages, Left-Right symmetry
cannot come down to a low energy for any choice of the parameter
space.
\section{Phenomenology}
\subsection{Proton Decay}
Having the mass scales and Higgs structure in hand we proceed in
this paper to discuss the issue of proton decay now. In all the
breaking chains that we have considered here, the quark lepton
unification is broken at the scale $M_{U}$ while the quark
antiquark unification is broken at the scale $M_{1}$. As a
result the leptoquark gauge bosons ($X_{\mu}$) will acquire mass
at the scale $M_{U}$ while the di-quark gauge bosons ($Y_{\mu}$) acquire mass
at the scale $M_{1}$. Under the group $G_{1}$ their
transformation properties are 
\begin{eqnarray}
X_{\mu} & \Longrightarrow & (6,1,-B,\bar{4})+(1,\bar{6},B,\bar{4})+\nonumber\\
        &    & (\bar{6},1,B,4)+(1,6,-B,4) \nonumber\\
Y_{\mu} & \Longrightarrow & (6,6,-2B,1)+(\bar{6},\bar{6},2B,1)\nonumber\\
where & & \nonumber\\
B & = & {1\over{2\sqrt 6}}\nonumber\\
\end{eqnarray}

Now $U(1)_{B}$ being an explicit local gauge symmetry of the
model, $X_{\mu}$ and $Y_{\mu}$ contains different " Baryon
Numbers " and hence cannot mix directly to form an $SU(16)$
invariant operator.

The mixing can be induced indirectly through the term $D_{\mu}\phi_{a}
D^{\mu}\phi_{b}$, where $D_{\mu}$ is the covariant derivative of
the $SU(16)$ invariant theory.$D_{\mu}\phi_{a}D^{\mu}\phi_{b}$
will contain a term $X_{\mu}\phi_{a}X^{\mu}\phi_{b}$. When $ \phi_{a}$
and $\phi_{b}$ acquires vacuum expectation value the mixing
between $ X_{\mu}$ and $Y^{\mu}$ occurs. But this can occur only at
the scale  $M_{6}$ hence the amplitude is suppressed by a factor
of ${O}{({M_{5}M_{6})}\over{(M^{2}_{1}M^{2}_{2}})}$.

To see how the gauge bosons couple to the Higgs fields we note
that all the gauge bosons at the $SU(16)$ level transform under
the $\bf{224}$ dimensional adjoint representation. We also note
the following tensor product at the $SU(16)$ level
\begin{equation}
{224 \times 224}=1+224+224+14175+10800+12376+12376\nonumber\\
\end{equation}

Being the product of two selfconjugate representations all the
terms in the right hand side are selfconjugate which couples to
only self conjugate representations. From the table1A that the
the Higgs field that carries Baryon Number is $1^{5}$. So the
only Higgs field which can induce a Baryon Number violating
effect is $1^{5}$ which is $\bf{4368}$ dimensional.

The only self conjugate combination made up with $1^{5}$s is
$<4368>$$<\bar{4368}>$ which again carries no baryon number hence
not giving rise to any baryon number violating process{\cite{opan}}.

To see the higgs field mediated proton decay at first we note
that the fermions are in the $\bf{16}$ dimensional fundamental
representation. To give mass to the fermions the coupling of the
form  ${\bar{\psi_L}}^{c}{\psi_{L}}{\phi}$  must exist. The
minimum dimensional Higgs field which can do the job is $\bf{120}$.
This field can give rise to Higgs mediated proton decay if $1^{6}$
breaks the Baryon Number due to the presence of the term
$<1^{6}>$$<1^{6}>$$<1^{2}>$$<1^{2}>$ in the Lagrangian.
In that case we can choose $\bf{136}$
to give mass to the fermions. 
In our choice $1^{5}$ breaks the
baryon number hence it does not couple to $\bf{120}$. Hence
there is no Higgs mediated proton decay.

\subsection{$N-\bar{N}$ oscillations}
Let us consider the $SU(16)$ level operator 
$<1^{5}>$$<1^{5}>$$<1^{5}>$$<16>$. This forms a singlet under
$SU(16)$ and hence allowed in the Lagrangian. This term give
rise to $\Delta B=3$ processes. If instead we choose $\bf{136}$ to
break the Lepton Number symmetry, then this process vanishes.

In the last subsecton we noted that if $1^{6}$ breaks the
Baryon Number symmetry then one has to choose $\bf 136$ to give
mass to the fermions here we note that then the term 
$<1^{14}1^{2}>$$<136>$$<136>$$<1^{6}>$$<1^{6}>$  will be be
allowed in the Lagrangian which may give rise to $\Delta B=3$
processes. As the term is of dimension five it will be
suppressed by $M_{U}$. 
With $1^{2}$ we can construct the $SU(16)$ level operator 
$<1^{5}>$$<1^{5}>$$<1^{4}>$$<1^{2}>$ which can break the Baryon Number
by two units and hence giving rise to  gauge boson mediated 
$N-{\bar{N}}$ oscillations. To see the Higgs field mediated
processes we note that if $\bf{120}$ dimensional Higgs field couples
to the fermions and $1^{6}$ breaks the Baryon number then 
the operator $<\bar{120}>$$<\bar{120}>$$<\bar{120}>$$<1^{6}>$
can give rise to Higgs field mediated N-{$\bar{N}$} oscillations.
\section{Conclusion}
In this paper we have seen that there exists one possible
breaking chain in a Grand Unified Theory based on the group
$SU(16)$ where a unification scale of the order of $10^{11}$
GeV is possible. There exists a very low energy scale  $(M_{B})$ which may
be almost anywhere between the unification scale and the
electroweak scale where completely ununified symmetry of quarks
and leptons may exist together with chiral color symmetry. The
scale $M_{B}$ comes lower when the separation between the scale
$M_{A}$ and the scale $M_{B}$ is increased. Qualitatively we
understand it in the following way. The beta function
coefficients can be looked into as the slope of the lines if
one plots the inverse coupling constants with respect to energy.
It can be easily checked that  as at the $SU(16)$ level all the 
fermions transform under
the fundamental representation of the group and in the other
levels they transform in a more complicated way under the various
groups in the intermediate stages all the groups cannot be
normalized in the same way. To compensate for the mismatch in
the normalisations the beta function coefficients has to be
multiplied by appropriate factors. Due to that the slope of the
curves representing the inverse couplings also gets multiplied
by the appropriate factors and the couplings get united earlier
giving rise to low energy unification.

We have also seen that this model satisfies the experimental
constraints coming from proton decay experiments in the sense
that proton decay is suppressed. We have shown that there exists
at least one choice of the Higgs sector where there is no Higgs
mediated proton decay either.

For some specific choice of the Higgs fields there may exist
interesting physical consequences like the $N-\bar{N}$
oscillation. There is also the possibility of having the sea-saw
mechanism to give Majorana mass to the neutrinos and this also
may have observable consequences.

Last but not the least we emphasize again that there exists very
rich low energy physics coming from this model hence keeping in
mind the forthcoming high-energy experiments at SSC,LHC and
other places this model is worthy of further investigation.

\section*{Acknowledgement}

It is a pleasure to thank Utpal Sarkar for introducing the
problem and also for a number of valuable discussions. The
author would also like to thank A. Joshipura and S.Rindani for
discussing some relevant points when this work was being
presented in a group seminar at P.R.L.

\section{Appendix}
\subsection{SU(16) Tensor Products}
\begin{eqnarray}
{16 \times{16}} & = & 120_{a}+136_{s}\nonumber\\
{\bar{16} \times{16}} & = & 1+255\nonumber\\
{16 \times 120} & = & 560_{a}+1360\nonumber\\
{\bar{120}} \times 120 & = & 1+255+14144\nonumber\\
{\bar{136}} \times 136 & = & 1+255+18240\nonumber\\
560_{a} \times 16 & = & 1820_{a}+7140\nonumber\\
1820_{a} \times 16 & = & 4368_{a}+24752\nonumber\\
\end{eqnarray}
\subsection{SU(16) Branching Rules}
\begin{eqnarray}
SU(16) & \Longrightarrow & SU(12) {\times} SU(4)\nonumber\\
16 & = & (12,1) +(1,4)\nonumber\\
136 & = & (78,1)+(12,4)+(1,10)\nonumber\\
120 & = & (66,1)+(12,4)+(1,6)\nonumber\\
255 & = & (143,1)+(12,\bar{4})+(\bar{12},4)+\nonumber\\
    &   &   (1,15)+(1,1)\nonumber\\
560 & = & (220,1)+(66,4)+(12,6)+\nonumber\\
    &   & (1,\bar{4})\nonumber\\
1820& = & (495,1)+(220,4)+(66,6)+\nonumber\\
    &   & (12,\bar{4})+(1,1)\nonumber\\
14144 & = & (1,1)+(1,35)+(12,\bar{4})+\nonumber\\
     &  & (12,\bar{20})+(\bar{12},4)+(\bar{12},20)+\nonumber\\
     &  & (\bar{66},6)+(66,\bar{6})+(143,1)+\nonumber\\
     &  & (143,15)+(70,\bar{4})+(\bar{780},4)+\nonumber\\
     &  & (4212,1)\nonumber\\
\end{eqnarray}
\subsection{SU(12) Tensor Products}
\begin{eqnarray}
{12 \times{12}} & = & 66_{a}+78_{s}\nonumber\\
{\bar{12} \times{12}} & = & 1+143\nonumber\\
{12 \times 66} & = & 220_{a}+572\nonumber\\
{\bar{78}} \times 78 & = & 1+143+5940\nonumber\\
{\bar{66}} \times 66 & = & 1+143+4212\nonumber\\
220_{a} \times 12 & = & 495+2145\nonumber\\
495_{a} \times 12 & = & 792+5148\nonumber\\
\end{eqnarray}
\subsection{SU(12) Branching Rules}
\begin{eqnarray}
SU(12) & \Longrightarrow & SU(6) {\times} SU(6){\times} U(1)\nonumber\\
12 & = & (6,1,-B) +(1,{\bar{6}},B)\nonumber\\
66 & = & (15,1,-2B)+(1,{\bar{15}},2B)+(6,{\bar{6}},0)\nonumber\\
78 & = & (21,1,-2B)+(1,{\bar{21}},2B)+(6,{\bar{6}},0)\nonumber\\
143 & = & (35,1,0)+(\bar{6},{\bar{6}},2B)+(6,6,-2B)+\nonumber\\
    &   & (1,1,0)+(1,35,0)\nonumber\\
220 & = & (20,1,-3B)+(1,{\bar{20}},3B)+(6,\bar{15},B)+\nonumber\\
    &   & (15,\bar{6},-B)\nonumber\\
495 & = & (15,1,-4B)+(20,{\bar{6}},-2B)+(15,\bar{15},0)+\nonumber\\
    &   & (6,\bar{20},2B)+(1,\bar{15},4B)\nonumber\\
792 & = & (\bar{6},1,-5B)+(15,{\bar{6}},-3B)+(20,\bar{15},-B)+\nonumber\\
    &   & (15,\bar{20},B)+(6,\bar{15},3B)+(1,6,5B)\nonumber\\
572 & = & (70,1,-3B)+(15,{\bar{6}},-B)+(6,\bar{15},B)+\nonumber\\
    &   & (21,\bar{6},-B)+(6,\bar{21},B)+(1,\bar{70},3B)\nonumber\\
4212 & = & (189,1,0)+(15,15,-4B)+(6,6,-2B)+\nonumber\\
    &   & (84,6,-2B)+(\bar{15},\bar{15},4B)+(1,35,0)+\nonumber\\
    &   & (1,189,0)+(\bar{6},\bar{84},2B)+(\bar{84},\bar{6},2B)+\nonumber\\
    &   & (6,84,-2B)+(1,1,0)+(35,1,0)+\nonumber\\
    &   & (35,35,0)+(\bar{6},\bar{6},2B)\nonumber\\
\end{eqnarray}
\subsection{$SU(6)$ Branching Rules}
\begin{eqnarray}
SU(6) & \Longrightarrow & SU(3) {\times} SU(2)\nonumber\\
6 & = & (3,2) \nonumber\\
15 & = & (6,1)+({\bar{3}},3) \nonumber\\
20 & = & (1,4)+(8,2) \nonumber\\
21 & = & (\bar{3},1)+(6,3) \nonumber\\
35 & = & (1,3)+(8,1)+(8,3)\nonumber\\
70 & = & (1,2)+(8,4)+(8,2)\nonumber\\
    &   & (10,2)\nonumber\\
\end{eqnarray}
\subsection{Normalization of $U(1)_{(B-L)}$ and $SU(2)^{q+l}_{R}$}
Consider chain3. Under the group $G_{6}$ the sixteen fermions
transform as 

($\bar{3}$,1,2,${1\over{\sqrt{24}}}$)+(3,2,1,$-{1\over{\sqrt{24}}}$)+
(1,1,2,$-{3\over{\sqrt{24}}}$)+(1,2,1,${3\over{\sqrt{24}}}$)

The $T^{3}_{2R}$ parts of the right handed $SU(2)$ group are to
be taken as$\pm{1\over{\sqrt{24}}}$ so as to get the correct U(1)
charges at the Standard Model level. $U(1)_{B-L}$ is normalized
to 2 while the $U(1)$ generator of the right handed $SU(2)$ is
normalized to $8\over{24}$ i,e $1\over{3}$. Taking this factor
of 6 in the relative normalization one can easily get the
familiar matching conditions of the $SO(10)$ model
\subsection{Anomaly Cancellation and Mass Scales}
$16$ dimensional fundamental representation of $SU(16)$ is not
anomaly free. To get the cancellation of anomaly one has to
introduce mirror fermions. But these fermions will not alter the
values of the mass scales obtained here. This is because in the two equations used for
$Sin^{2}(\theta_{w})$ and $1-{8\over3}{\alpha\over\alpha_{s}}$ respectively
the fermion contributions to the beta function coefficients cancel.

\vskip 1in

\par\vfill\eject

\end{document}